\documentclass{moriond}

\def\be{\begin{equation}}
\def\ee{\end{equation}}
\def\bea{\begin{eqnarray}}
\def\eea{\end{eqnarray}}

\usepackage{ulem}
\usepackage[dvipsnames]{xcolor}

\usepackage{physics}
\usepackage{siunitx}

\usepackage{cleveref}

\newcommand{\Photo}{}

\begin{document}
\vspace*{4cm}
\title{Impact of blending on weak lensing measurements with the Legacy Survey of Space and Time}

\author{ M. Ramel, C. Doux, M. Kuna }

\address{Université Grenoble Alpes, CNRS, LPSC-IN2P3, 38000 Grenoble, France}

\maketitle\abstracts{
Upcoming deep optical surveys, such as the Vera C. Rubin Observatory Legacy Survey of Space and Time (LSST), will scan the sky to unprecedented depths, detecting billions of galaxies. However, this amount of detections will lead to the apparent superposition of galaxies in the images, a phenomenon known as blending, that can affect the accurate measurement of individual galaxy properties. In particular, galaxy shapes play a crucial role in estimating the masses of large-scale structures, such as galaxy clusters, through weak gravitational lensing.
This proceeding introduces a new catalog matching algorithm, \texttt{friendly}, designed for detecting and characterizing blends in simulated LSST data for the Dark Energy Science Collaboration (DESC) Data Challenge 2. The aim of this algorithm is to combine several matching procedures, as well as a probabilistic method to quantify blended systems.
By removing the resulted 27\% of galaxies affected by blending from the dataset, we demonstrate that the amplitude of the excess surface mass density $\Delta\Sigma$ weak lensing profile — potentially biased low due to blending — may be partially corrected.}


\section{Introduction}\label{sec:intro}

\subsection{Cosmology with galaxy clusters}\label{subsec:cosmo}

Galaxy clusters, as the largest structures in the Universe, are the main tracers of the highest peaks in the total matter density field. Since the history and evolution of structure formation are sensitive to the number and masses of matter overdensities, clusters of galaxies are important probes, used to infer cosmological parameters. 

However, masses of galaxy clusters are not directly measurable and have to be determined through indirect effects such as weak gravitational lensing. Gravitational lensing occurs when light rays coming from background galaxies are bent by the high masses of large foreground structures, such as galaxy clusters \cite{Hoekstra2008weaklensing}. This effect results in the distortion and magnification of background galaxies. The measurement of the distorted shapes $\epsilon^{\rm obs}$ of these latter allows to determine an estimator of the excess surface mass density $\widehat{\Delta\Sigma}$ as function of the distance to the center of the structure, $R$, given as:
\begin{equation}
    \widehat{\Delta\Sigma}(R) = \expval{ \Sigma_{\rm crit}(z_{\rm gal}, z_l) \, \epsilon_+^{\rm obs} } (R),
\end{equation}
where ${\Sigma_{\rm crit}}$ is a geometrical term depending on the redshifts of the lens $z_l$ and of the source galaxies $z_{\rm{gal}}$, and $\epsilon_+^{\rm{obs}}$ are the tangential ellipticities of the sources. By fitting the $\Delta\Sigma$ profiles with a Navarro-Frenk-White halo mass profile model, estimated projected masses of galaxy clusters can be recovered \cite{Navarro1996NFW}.


\subsection{The Vera C. Rubin Observatory}\label{subsec:lsst}

The next stage of future deep optical surveys will bring a large amount of data, including measurements of shapes and redshifts of galaxies, used to estimate galaxy cluster masses through weak gravitational lensing. From 2025 onwards, the Legacy Survey of Space and Time (LSST)~\cite{{Ivezic2019LSST}} survey will be conducted by the Vera C. Rubin Observatory, currently in construction in northern Chile. During the ten years of observations of a \num{18000} squared degrees footprint, about 10 billions of galaxies up to a magnitude of 27.5 in $i$-band will be observed.


\subsection{Blending}\label{subsec:blending}

Due to the high depth of observation and the atmosphere-limited resolution of future optical ground-based surveys such as LSST, galaxies may overlap along the line of sight and on images. This observational effect, known as blending, will impact the galaxy property measurements.

Two types of blends can be distinguished. We define \textit{recognized blends} as systems where two or more galaxies significantly overlap but are still detected as individual objects, depending on the detection pipeline. With LSST, the fraction of recognized blends is estimated to be around 40~\% \cite{Dawson2015BlendingShapes}. Conversely, \textit{unrecognized blends} occur when galaxies are so overlapped that they cannot be identified individually. These are the most challenging cases as they won't be detectable in future LSST data. Around 20~\% of galaxies may be part of unrecognized blends in LSST-like surveys \cite{Troxel2023BlendProportion}. This will consequently impact measurements of individual galaxy properties such as shapes \cite{Dawson2015BlendingShapes} or redshifts, used for determining masses of galaxy clusters through weak lensing.


\section{Matching procedure}\label{sec:matching}

To study the impact of blending on future LSST weak lensing data, we must first identify blended systems. To do so, we use DESC \cite{Mandelbaum2018DESC} simulations to compare LSST-like simulated catalogs with truth or reference data.
\subsection{Simulated catalogs}\label{subsec:catalogs}

The reference catalog, \texttt{cosmoDC2} \cite{Korytov2019CosmoDC2}, spans 440 square degrees of the sky, constructed from a dark matter $N$-body simulation and serving as the starting point for the DESC Data Challenge~2 \cite{Abolfathi2021DC2}. Each galaxy in the catalog is described by properties such as positions, true redshifts, intrinsic ellipticities or shears.

From the \texttt{cosmoDC2} catalog, 300 squared degrees have been simulated to create realistic images of the sky and processed by the Rubin Science Pipeline \cite{Jenness2022Pipeline}. A second catalog, \texttt{DC2object}, contains the measured quantities of detected objects from these images \cite{Abolfathi2021DC2}.


\subsection{\texttt{friendly} matching algorithm}\label{subsec:friendly}

We developed a matching algorithm that exploits position, flux and shape information to progressively refine an initial crude matching and better characterize blends in DC2 simulated data.

The first step uses a Friends-of-Friends (FoF) algorithm~\footnote{https://github.com/yymao/FoFCatalogMatching} to identify groups of nearby detected objects from \texttt{DC2object} and simulated galaxies from \texttt{cosmoDC2}, based on their angular distances and using a relatively large linking length of 2''. 

We then use shape information to refine these groups. To do so, we associate an ellipse to each galaxy (object) using their true (measured) moments, and perform an ellipse overlap test~\footnote{Developed by collorator Shuang Liang: https://github.com/LSSTDESC/Cluster\_Blending/
} to remove FoF links of non-overlapping pairs, and break FoF groups into smaller ones. This matching procedure is currently implemented in the \texttt{friendly} matching algorithm~\footnote{https://github.com/LSSTDESC/friendly/tree/FoF}.

Once the groups have been formed, we can easily identify blended systems, depending on the number of detected objects in comparison with nearby truth galaxies in each group. In this paper, we will label $n-m$ systems those composed of $n$ galaxies and $m$ objects. We note immediately that the characteristics of these groups depend on the various cuts (e.g. on magnitude) applied to the two catalogs.


\subsection{Blending entropy}\label{subsec:entropy}

Blending results in some detected objects being improperly matched to their truth galaxies, leading to inaccurate measurements of their shapes and redshifts. We propose thus to quantify blending as a matching \textit{ambiguity}. For this purpose, we define the relative probability of matching, computed for each detected object with respect to the truth galaxies of its FoF group, as a function of their positions, fluxes and shapes.

To characterize the level of ambiguity, we then use the (normalized) matching probabilities to compute the \emph{blending entropy} for each detected object, defined as: 
\begin{equation}
    S_b = -\sum_i{p_i \log{p_i}},
\end{equation}
where $p_i$ is the probability of matching the $i$-th galaxy of a given object's group. Objects with high blending entropies are associated with highly blended groups, while by definition, the blending entropy of well-matched systems equals 0.

\begin{figure}
\centerline{\includegraphics[scale=0.9]{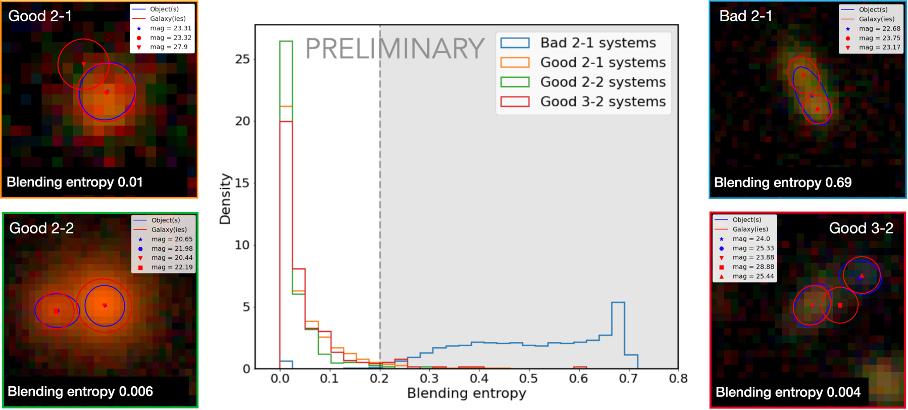}}
\caption{Blending entropy distributions of detected objects in \textit{bad} $2-1$ (in blue) and \textit{good} $2-1$ (in orange), $2-2$ (in green), and $3-2$ (in red) systems. Example of such systems are illustrated with LSST-DESC simulated images. A 0.2 cut would allow to isolate highly blended objects, as shown with the grey area.}
\label{fig:Entropy}
\end{figure}

To illustrate the discriminative power of this new quantity, we compute its distribution for specific $2-1$ systems, either highly blended (\textit{bad}) or well-identified by the LSST pipeline (\textit{good}), as shown in \cref{fig:Entropy}. In particular, we find that applying a blending entropy cut of approximately 0.2 effectively isolates objects involved in \textit{bad} blends while conserving perfectly and relatively well-matched objects for future studies.

\section{Impact of blending on $\Delta\Sigma$ profiles}\label{sec:wl}

\begin{figure}
\centerline{\includegraphics[scale=0.7]{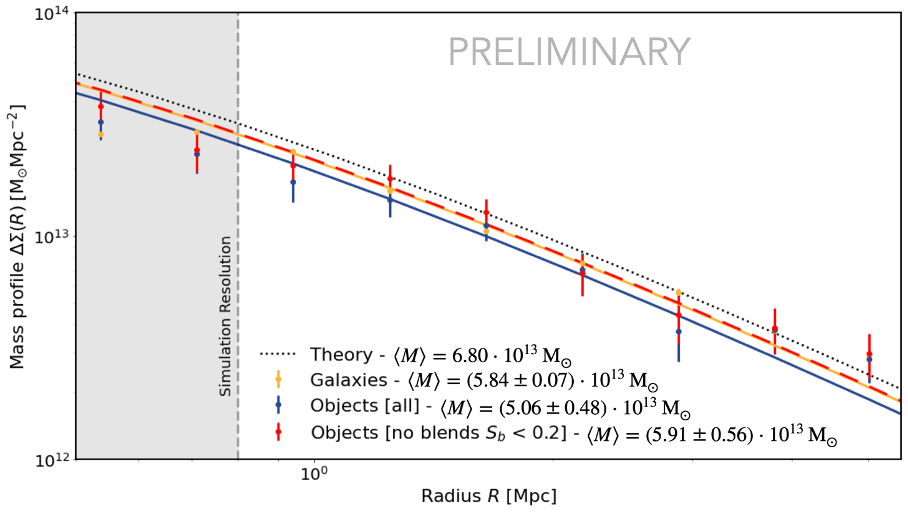}}
\caption{Impact of blending on stacked $\Delta \Sigma$ profile. The yellow profile corresponds to the stacked profile measured using the shapes and redshifts of background galaxies from \texttt{cosmoDC2}. The red (respectively blue) profile is measured with (respectively without) a blending entropy cut on \texttt{DC2object} observation data. The fiducial profile is plotted in black.}
\label{fig:DeltaSigma}
\end{figure}

The objective of this work is to study the impact of blending on cluster stacked $\Delta\Sigma$ profiles for the future LSST weak lensing data analysis. Preliminary results are shown in \cref{fig:DeltaSigma}. 

In this proceeding, we compare the stacked lensing profile~\footnote{computed with CLMM: https://github.com/LSSTDESC/CLMM} of all detected objects from \texttt{DC2object} with the one obtained after removing objects impacted by blending. To achieve this, we apply a blending entropy cut of 0.2 on the detected objects, as discussed in \cref{subsec:entropy}, effectively isolating and mitigating the effects of problematic blends. This results in a 27~\% suppression of objects. For comparison purpose with simulated truth, the stacked $\Delta\Sigma$ profile of the lensed galaxies from \texttt{cosmoDC2} is also plotted. The grey area of \cref{fig:DeltaSigma} corresponds to the maximum radius within which the resolution of the weak lensing simulation is not precise enough to be used for estimating the masses of galaxy clusters.

As a preliminary result, we observe that removing blends may shift the lensing profile upwards, bringing it closer to the reference profile measured by \texttt{cosmoDC2} data. This result suggests that blending can indeed influence $\Delta \Sigma$ lensing profiles, leading to a reduced amplitude of the lensing signal and thus underestimates of galaxy cluster masses.

\section{Conclusion}\label{sec:conclusion}

In this study, we introduce a new catalog matching algorithm named \texttt{friendly}, designed for identifying and characterizing blended systems in forthcoming LSST weak lensing data. By defining the relative probability of object-truth matches, combining positions, shapes, and magnitudes information, we assess the level of matching ambiguity using the blending entropy $S_b$. Upon removing highly blended objects with a $S_b$ cut-off of 0.2, we observe that blending may introduce a bias in the amplitude of the stacked $\Delta\Sigma$ lensing profile. This leads to a weaker signal and underestimates in the masses of galaxy clusters. Further work will be done in the future to propagate this study to cosmological parameters.


\end{document}